\documentclass[10pt, a4paper]{article}

\usepackage{lrec-coling2024} 

\usepackage[TU]{fontenc}

\usepackage{amsmath}
\usepackage[capitalize]{cleveref}
\usepackage{graphicx}
\usepackage{subcaption}
\usepackage{multirow}

\usepackage{hyperref}
\usepackage{academicons}
\usepackage{xcolor}

\newcommand{\orcid}[1]{\href{https://orcid.org/#1}{\textcolor[HTML]{A6CE39}{\aiOrcid}}}

\title{Multi-word Term Embeddings Improve Lexical Product Retrieval}

\name{Fedor Krasnov\textsuperscript{1}, Viktor Shcherbakov\textsuperscript{2}\textsuperscript{3}} 

\address{\textsuperscript{1} Research Center of Wildberries SK LLC based on the Skolkovo Innovation Center, \\ \textsuperscript{2}University of Geneva, \textsuperscript{3}University of Lausanne \\
         krasnov.fedor2@wb.ru, Viktor.Shcherbakov@unil.ch}

\abstract{
Product search is uniquely different from search for documents, Internet resources or vacancies, therefore it requires the development of specialized search systems. The present work describes the H1 embdedding model, designed for an offline term indexing of product descriptions at e-commerce platforms. The model is compared to other state-of-the-art (SoTA) embedding models within a framework of hybrid product search system that incorporates the advantages of lexical methods for product retrieval and semantic embedding-based methods. We propose an approach to building semantically rich term vocabularies for search indexes. Compared to other production semantic models, H1 paired with the proposed approach stands out due to its ability to process multi-word product terms as one token. As an example, for search queries "new balance shoes",  "gloria jeans kids wear" brand entity will be represented as one token - "new balance", "gloria jeans". This results in an increased precision of the system without affecting the recall. The hybrid search system with proposed model scores mAP@12 = 56.1\% and R@1k = 86.6\% on the WANDS public dataset, beating other SoTA analogues.\\ \newline  }

\begin{document}

\maketitleabstract

\section{Introduction}

Product search systems are required to operate with both low latency and high recall, since they scan the whole product catalog of billions of items. Common product search methods initially used lexical search models. These models calculate the relevance metric based on heuristics that measure exact word match between the search query and textual product representations. Lexical search models such as BM25 \citep{BM25} have been relevant for decades, and are still widely used today. The recent alternatives, neural extraction methods, demonstrate increased search effectiveness metrics, but also possess their own flaws \citep{FAERY, DFGAS, RECOM, CAA}. Naturally, the research gravitates towards the hybridization of the two approaches, combining the advantages of each.

The disadvantages of lexical models are well-researched: (E1) a possible mismatch between query and document vocabularies \citep{OOV1, OOV2} leads to search recall degradation; (E2) lack of semantic understanding of queries and documents \citep{OOV3} decreases search precision. These described limitations result in failures to retrieve relevant documents using lexical methods for information retrieval. To resolve these issues a number of extensions to the lexical model have been introduced in the past decades, including, but not limited to: query expansion \citep{QR1, QR2, QR3, QR4}, document expansion \citep{DR1, DR2, DR3}, term dependencies model \citep{QD1, QD2}, topic modeling \citep{TM1, TM2}, machine translation models for information retrieval \citep{MLIR1, MLIR2}. Despite mentioned advances, the research in lexical models for information retrieval progresses relatively slowly, since the majority of these methods work with discrete, sparse lexical representations and inevitably inherit their limitations.

\begin{figure*}[ht]
    \centering
    \includegraphics[width=\linewidth]{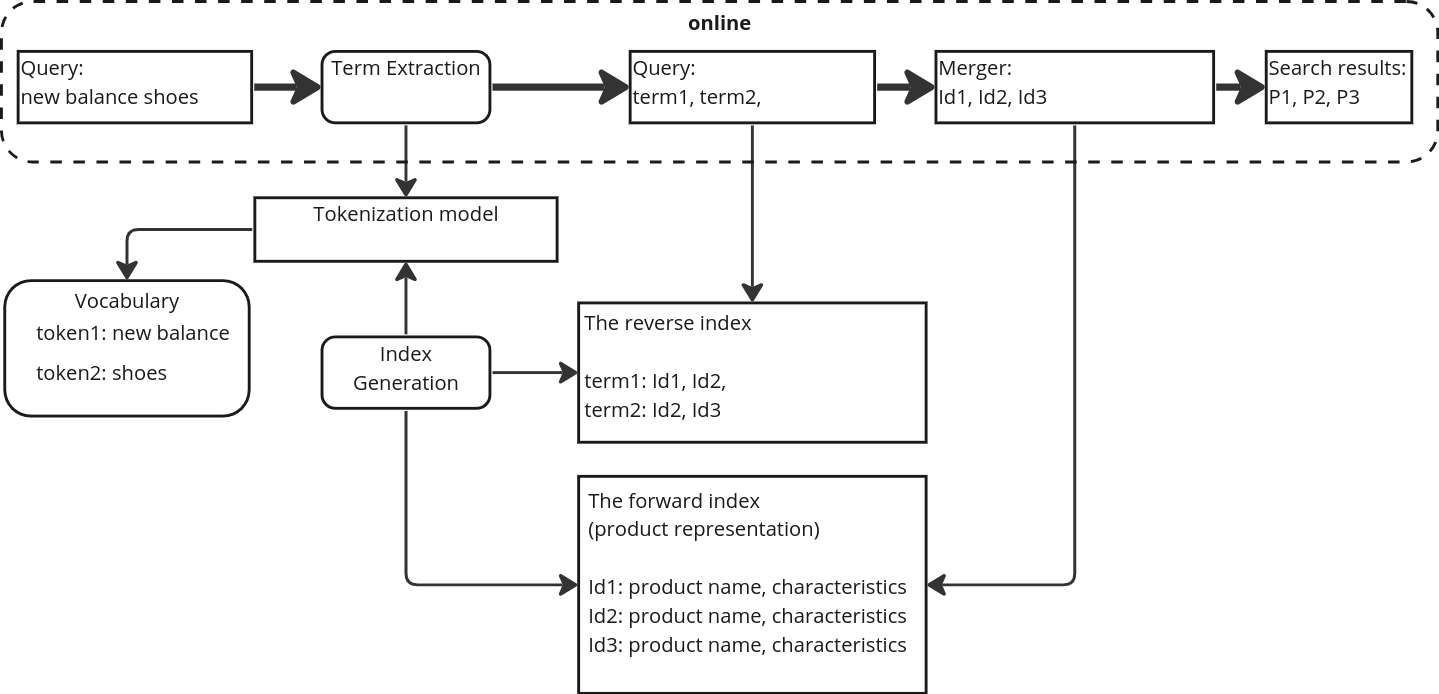}
    \caption{The indexes and a tokenization model built during offline information extraction are used in online setting to respond to queries with low latency. The quality of built index is detrimental to the performance of the search system.}
    \label{fig:role}
\end{figure*}

With the development of representation learning in information retrieval, semantic search models at the offline information extraction stage of the search have seen an increased research interest in recent years. During this stage the indexes are built for matching queries with the documents. The Figure \ref{fig:role} schematically describes an example product search system that uses indexes built during the information extraction stage for fast responses to queries. 

Starting in 2013, the improvement of word embeddings \citep{EMB1, EMB2, EMB3} has led to a number of studies using embeddings for the extraction stage \citep{EBR1, EBR2, EBR3}. Unlike discrete lexical representation, word embeddings offer a continuous representation that can help with the problem of query and document vocabularies mismatch to some extent. After 2016, a spike of research attention to the application of deep learning methods to the information extraction stage is seen \citep{IR1, IR2}. These methods are applied either for improving document representation within the framework of the traditional paradigm of discrete lexical representation \citep{TF1, TF2, TF3}, or directly for forming novel semantic search models within the sparse/dense representation paradigm \citep{DIR1, DIR2, ColBERT, DIR4}. 

While closely related to document information retrieval, the product search problem is uniquely different in a few aspects:
\begin{itemize}
\item Ranking mechanisms based on weighing textual features (TF/IDF, BM25) differ in product search. For example, the token frequency in the product title does not affect the query relevancy.
\item Products are multimodal. A product page includes a title, description, characteristics, images, videos, etc. The search system can take into account multiple modalities of a page.
\item Search queries are motivated by an interest in purchasing a product. Customer behavior differs significantly from vacancy search or Internet resource search behavior.
\item Product search effectiveness is evaluated on a modality-wise basis.  
\end{itemize}

The primary research question of the present paper is to evaluate the impact of the semantic model and tokenization architectures on offline metrics of a hybrid product search system.

In the following sections, we describe in detail the research methodology, conducted experiments and conclusions. 

\section{Related work}

\subsection{Neural Information Retrieval}

Similar to document information retrieval trends, the development of product search systems has transitioned from lexical retrieval methods to neural retrieval methods \citep{TAOBAO, WALMART, SEMSEARCH}. DSSM \citep{DSSM}, being one of the most popular neural network architectures, is based on a Dual Encoder paradigm \citep{Gillick2018, Yang2020, Karpukhin2020}. The two independent “towers” of encoders—one for search queries and the other for product representation—embed queries and products into a shared space of fixed dimensionality. The shared space is used for similarity search \citep{Vanderkam2013, Johnson2021} to retrieve products that are relevant to a search query. Thus far, the most promising results have been achieved by using the BERT model in a Dual Encoder architecture \citep{Tower-BERT, BERT-Siamese, TwinBERT}. The general operating principle of these models is described in \cref{eq:pool_q,eq:pool_p,eq:sim_bert}. 

\begin{align}
    \overrightarrow{q} = AvgPool \left [ BERT_{\theta}^l(q) \right] 
    \label{eq:pool_q} \\
    \overrightarrow{p} = AvgPool \left [ BERT_{\theta}^r(p) \right] 
    \label{eq:pool_p} \\
    s_{BERT} (\overrightarrow{q}, \overrightarrow{p}) = \overrightarrow{q}^T \cdot \overrightarrow{p}
    \label{eq:sim_bert}
\end{align}

Where $BERT^t_{\theta}$ and $BERT^r_{\theta}$ are the “left” and “right” encoders, respectively, transforming texts $q$ and $p$ into a shared space $\theta$. The similarity function $s_{BERT}(\cdot, \cdot)$ is implemented with a scalar product of $\overrightarrow{q}$ and $\overrightarrow{p}$. The bottleneck in this architecture lies in the averaging of the token vectors.

The ColBERT \citep{ColBERT} model represents a particular variant of the Dual Encoder architecture, termed a Single Encoder. Models based on this architecture use the same encoder for both queries and products. However, the novelty of ColBERT lies in computing the similarity scores token-wise, instead of comparing the mean vectors. Given a search query $q$ comprising $m$ tokens and a product $p$ comprising $n$ tokens, the similarity function $s_{ColBERT}(\cdot, \cdot)$ is:

\begin{equation}
    s_{ColBERT}(q_{1:m}, p_{1:n}) = \sum_1^m \max_{1..n}\left (\overrightarrow{q}^T_{1:m} \cdot \overrightarrow{p}_{1:n} \right )
    \label{eq:sim_colbert}
\end{equation}

The sum over maximum similarity scores for each token of a query in \cref{eq:sim_colbert} implies that $n \cdot m$ scalar products need to be calculated, compared to one scalar product in $s_{BERT}(\cdot, \cdot)$.

\subsection{Hybridization}

It is accepted to understand hybridization as mixing the lexical and neural methods of information retrieval within one product search system. Hybridization can be applied at different stages of the search. For instance, the authors of the study \citet{SEMSEARCH} combined the search results of several distinct models based on lexical, behavioral, and semantic methods. Another hybridization principle was applied in the study \citet{CLEAR}—the lexical method was the primary retrieval mechanism, while a semantic model was trained to correct the mistakes of the lexical model.

\subsection{Tokenization}

The progress in tokenization methods has led to significant improvements in the offline metrics of natural language processing models \citep{SP, BPE}. The BPE (Byte-Pair-Encoding) tokenization method was originally introduced as a data compression method \citep{Gage1994}. In constructing the BPE tokenizer, the initial vocabulary is sequentially extended until the preset limit is reached. The primary goal of applying BPE to natural text is to split words into commonly occurring subwords. Usually, little care is given to the actual semantics of the final tokens. However, unlike most applications, where semantic information can be represented by the combination of tokens, information retrieval often requires semantically rich tokens in order to use them as terms to construct effective search indexes, see Fig \ref{fig:role}.

The later proposed alternative, the unigram tokenization method \citep{UNI}, demonstrates the opposite approach—the vocabulary size is sequentially pruned by removing rare tokens that can be replaced by common tokens. The unigram method was primarily introduced to provide multiple possible tokenizations for a given text with the use of a unigram language model. During vocabulary construction, both methods aim to minimize the length of text encoded in tokens and, in practice, produce similar tokenizations.

\section{Methodology}

\subsection{H1}

\begin{figure*}[t]
    \centering
    \includegraphics[width=0.8\linewidth]{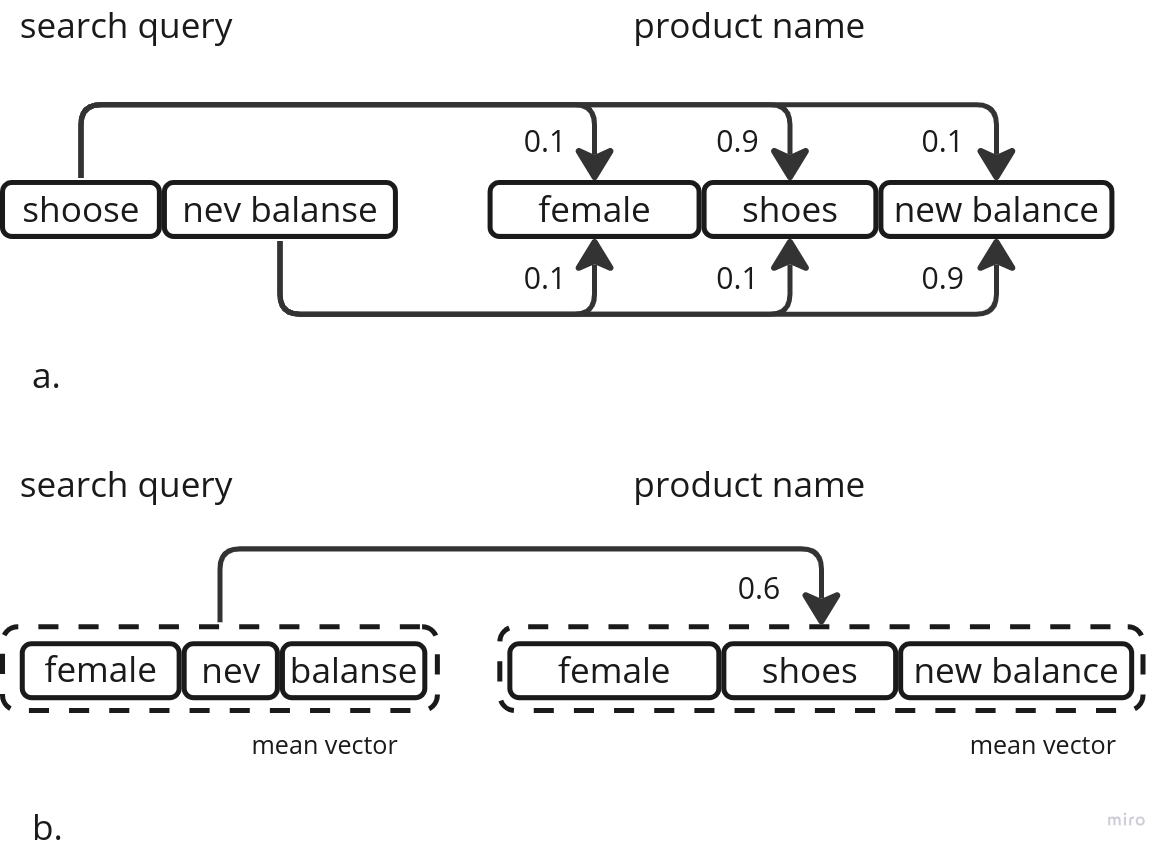}
    \caption{Token handling principle of H1 (a), compared to that of a FastText (b). H1 attributes scores to each pair of query and document tokens, while the FastText-based system compares the mean vector representations.}
    \label{fig:token-handling}
\end{figure*}

The H1 semantic model draws significant inspiration from ColBERT but architecturally simplified. It processes both queries and documents by tokenizing them and then passing them through a BERT-based Dual Encoder. The resulting embeddings are evaluated using the $s_{ColBERT}(\cdot, \cdot)$ similarity function. We explore the impact of different tokenization techniques in Ablation Study Section \ref{sec:ablation}. A distinctive aspect of H1 is its approach to token-level lexical hybridization, where we enhance the tokenizer's vocabulary with semantically rich terms to improve the semantic independence of standalone terms. The Experiments Section \ref{sec:experiments} provides a comprehensive analysis of the H1 system's application in a product retrieval task. For this task specifically, we augmented the tokenizer's vocabulary with a carefully selected list of brand names. 

The rationale behind incorporating brand names into the vocabulary is rooted in understanding user search behavior, particularly when it comes to specific brands. For instance, when a customer searches for "new balance shoes", their intent is not to explore products related to the terms "new" and "balance" independently. Instead, they are looking for items specifically associated with the "New Balance" brand. However, these customers may still be open to considering various types of "shoes". 

H1 model is optimized on positive and negative product-query pairs using the following loss function:

\begin{equation}
    L_{H1} = \left[ \gamma - s_{\theta}(q_{1:m}, p_{1:n}^{+}) + s_{\theta}(q_{1:m},p_{1:n}^{-}) \right]_{+0}
    \label{ref:loss}
\end{equation}

Where $\gamma$ is a threshold and $s_{\theta}$ is a similarity relation parametrized by $\theta$, applied to an $m$-token query $q$ with a positive $p^{+}$ and a negative $p^{-}$ product description example. Negative examples are sampled by selecting a random product from the current batch. The square brackets around the equation, $[]_{+0}$, denote that negative values are set to 0.

\subsection{Evaluation}

\label{sec:evaluation}

Neural retrieval methods, given their computational intensity, are impractical for online product searches within catalogs containing billions of items. Instead, their utility shines in building indexes for product descriptions, as schematically outlined in the example in Figure \ref{fig:role}. The actual neural encoder is never utilized to generate the embeddings for user queries. Our evaluation methodology mirrors these practical limitations, ensuring that our approaches are both realistic and aligned with the constraints of large-scale product retrieval systems.

For the query encoder $E_{\theta}^q$, the product encoder $E_{\theta}^p$, the similarity measure $s$, and the tokenization method $T$, the evaluation procedure employed in the experiments (Section \ref{sec:experiments}) is as follows:

\begin{enumerate}

\item The vocabulary of query tokens $V_q$ is collected using $T$.
\item For every token $t_i$ from the vocabulary $V_q$, its embedding $e_i^q = E_{\theta}^q[t_i]$ is produced.
\item The embeddings for the tokens in every product description $p_{1:n}^j$ are computed as: $$ (e_{j, k})_{k=1}^n = E_{\theta}^p[T(p_{1:n}^j)] $$
\item An index that maps every query term to relevant products is built using query tokens as terms: $$ I(t_i) = \{ p_{1:n}^j ~ | ~ s(e_i^q, e_{j, 1:n}^p ) > \gamma \} $$ where $\gamma$ is a relevancy threshold.
\item For a query $q_{1:m}$ with tokens $$T(q) = (t’_1, ~ ...  ~, t’_m)$$ a list of all relevant products according to the index $I$, $$R = I(t'_1) | ~ …  ~ | I(t'_m)$$ is collected, and the metric is computed on $R$, sorted with respect to the similarity of relevant products to the query.

\end{enumerate}

The described evaluation approach mimics the product search implemented with a simple term index-based hybrid search system. This system combines the efficiency of fast lexical term lookup in an index for high precision, with the computation of similarity scores on only a subset of all product descriptions, ensuring low latency responses. The performance of the system is entirely dependent on the similarity measure $s$ within embedding space defined by semantic model of choice.

The offline metrics for product search differ from those of document information retrieval. The objective of product search is to identify several, or ideally, all products relevant to the query, including identical items. This requirement stems from the customer's need to compare prices for identical products. Hence, the formulas for recall and precision are adapted to include an equivalence relation $M$. Precision metrics for product search are defined as follows, with recall metrics being similarly formulated.

\begin{align}
    P @ k = \frac{1}{|Q|} \sum_{q \in Q} \frac{\left|M(p_q^r @ k, p_q^g)\right|}{k}
    \label{eq:precision} \\
    mAP @ k = \frac{1}{k} \sum_{i=1}^k P @ i
    \label{eq:m_ap} 
\end{align}

\begin{description}
    \item[$Q$] -- the set of all search queries.
    \item[$p_q^g$] -- all ground truth products for query $q$.
    \item[$p_q^r$] -- the retrieved products for query $q$ at rank $k$.
    \item[$M(A, B)$] -- the set of products in $A$ that are equivalent to any of the products in $B$.
\end{description}

\section{Experiments}

\begin{figure*}[ht]
    \centering
    \begin{subfigure}[b]{\textwidth}
        \centering
        \includegraphics[width=\textwidth]{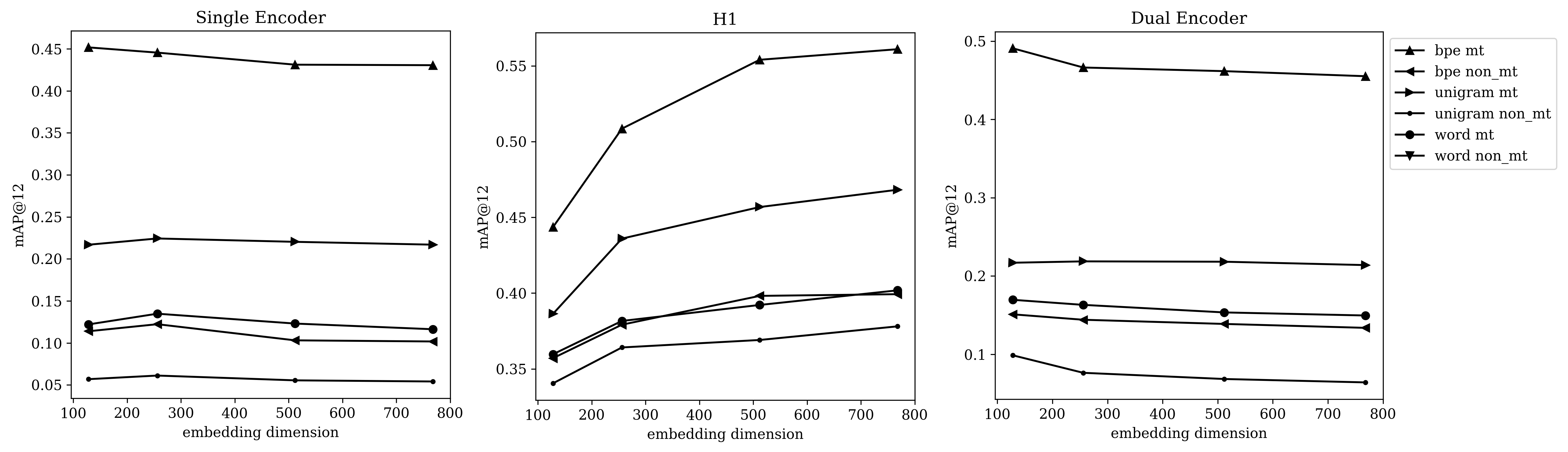}
    \end{subfigure}

    \begin{subfigure}[b]{\textwidth}
        \centering
        \includegraphics[width=\textwidth]{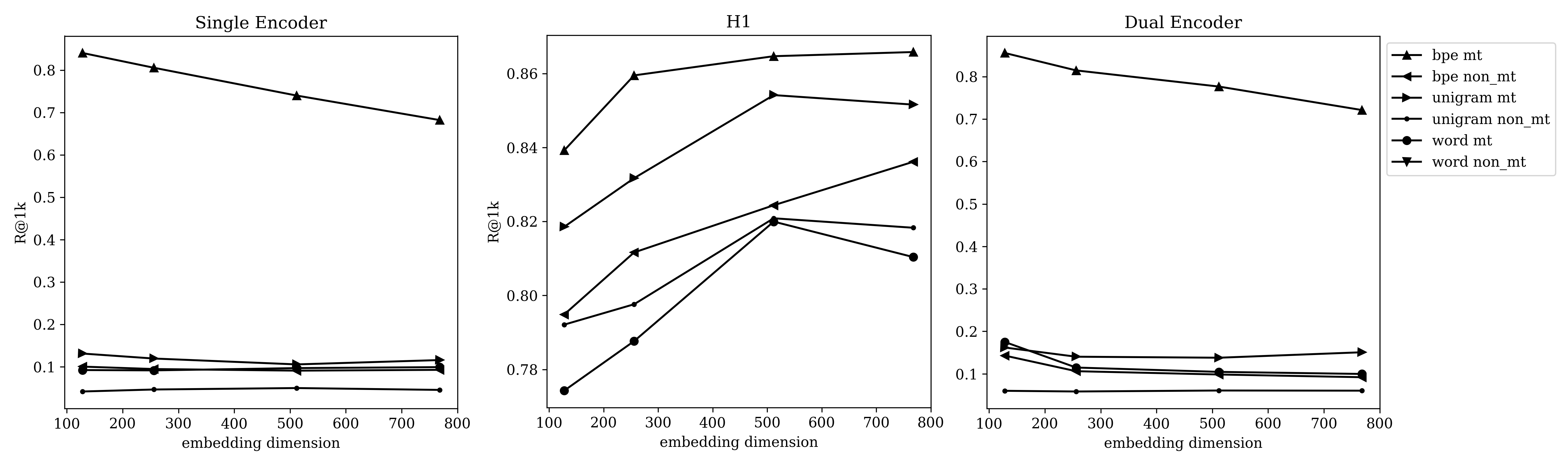}
    \end{subfigure}

    \caption{Ablation study results over tokenization methods and model architectures.}
    \label{fig:ablation}
\end{figure*}

\label{sec:experiments}

We evaluated the proposed H1 model against several existing information retrieval models, specifically TCT-ColBERT \citep{TCT-ColBERT}, Single Encoder (SE) \citep{SEMSEARCH}, and Dual Encoder (DE) \citep{DSSM}. Additionally, we experimented with three tokenization methods: Byte Pair Encoding (BPE), unigram, and word tokenizations. For each tokenization method, we proposed two variations: one enriched with a predefined set of brand names as special tokens (referred to as multi-token or {\it mt} variations), and a standard version without added brand names (non-multi-token or {\it non-mt } variations).

We employed the SentencePiece library for all tokenization tasks, configuring it with the {\it split\_by\_whitespace=False} option to ensure multi-word brand names could be incorporated as special tokens.

Following the evaluation methodology outlined in Section \ref{sec:evaluation}, we calculated the metrics mean Average Precision at 12 items ($mAP@12$) and Recall at 1000 items ($R@1k$) for H1, SE, DE models combined with every tokenization method described earlier. Two products are considered to be equivalent if they share the same title.

We compare the performance of the best combination of the model type and tokenization method against ColBERT implemented by Terrier \citep{PyTerrier} and trained with Tight Coupling Teachers method \citep{TCT-ColBERT}. 

\subsection{Dataset}

Our data source is the publicly available WANDS dataset, chosen for its suitability in objectively benchmarking retrieval systems in the context of e-commerce. The dataset's key characteristics are as follows:

\begin{itemize}
\item 42,994 product candidates,
\item 480 queries,
\item 233,448 relevancy scores for query-product pairings.
\end{itemize}

The relevancy of query-product pairs in the WANDS dataset is annotated with three levels: fully relevant (Exact), partially relevant (Partial), and Irrelevant. For the purposes of training our models, we utilized only two labels: Exact (labeled as 1) and Irrelevant (labeled as -1), with class balancing implemented prior to training.

\subsection{Ablation study}

\label{sec:ablation}

First, we ablate over the tokenization method and model hyperparameter (embedding dimensions) for each of the model types: H1, SE, DE. For ColBERT model, the pretrained version was used, so it was not included in the ablation study. The results of the experiment are shown in Fig \ref{fig:ablation}. The best results, $R@1k = 86.6\%$ and $mAP@12 = 56.1\%$, were achieved by the combination of H1 model with 768 embedding dimensions and BPE tokenization with brand names added.

We note that for both BPE and unigram tokenizations, the variation with brand names added (mt) produces consistently better results for any model with any embedding dimensionality.

\subsection{Best models comparison}

\begin{table}[ht]
\centering
\begin{tabular}{|l|l|l|l|}
\hline
Model & Threshold & Precision & Recall \\ \hline
\multirow{4}{*}{ColBERT} & 12 & \textbf{41\%} & 26\% \\ \cline{2-4}
& 128 & 21\% & 61\% \\ \cline{2-4}
& 512 & 9\% & 78\% \\ \cline{2-4}
& 1024 & 5\% & \textbf{84\%} \\ \hline
\end{tabular}
\caption{The ColBERT results on WANDS dataset.}
\label{tab:colbert}
\end{table}

\label{sec:compare}

The Table \ref{tab:colbert} presents the results of the evaluation of the ColBERT model on the WANDS dataset with varying thresholds. The H1 model demonstrates better results, especially for Precision at 12 items.

To further demonstrate the superiority of the H1 model, we compare H1, SE, DE, and ColBERT models with the best hyperparameters seen in the Ablation Study Section \ref{sec:ablation} on a single query with multiple thresholds $k$.

\begin{figure}[t]
    \centering
    \includegraphics[width=\linewidth]{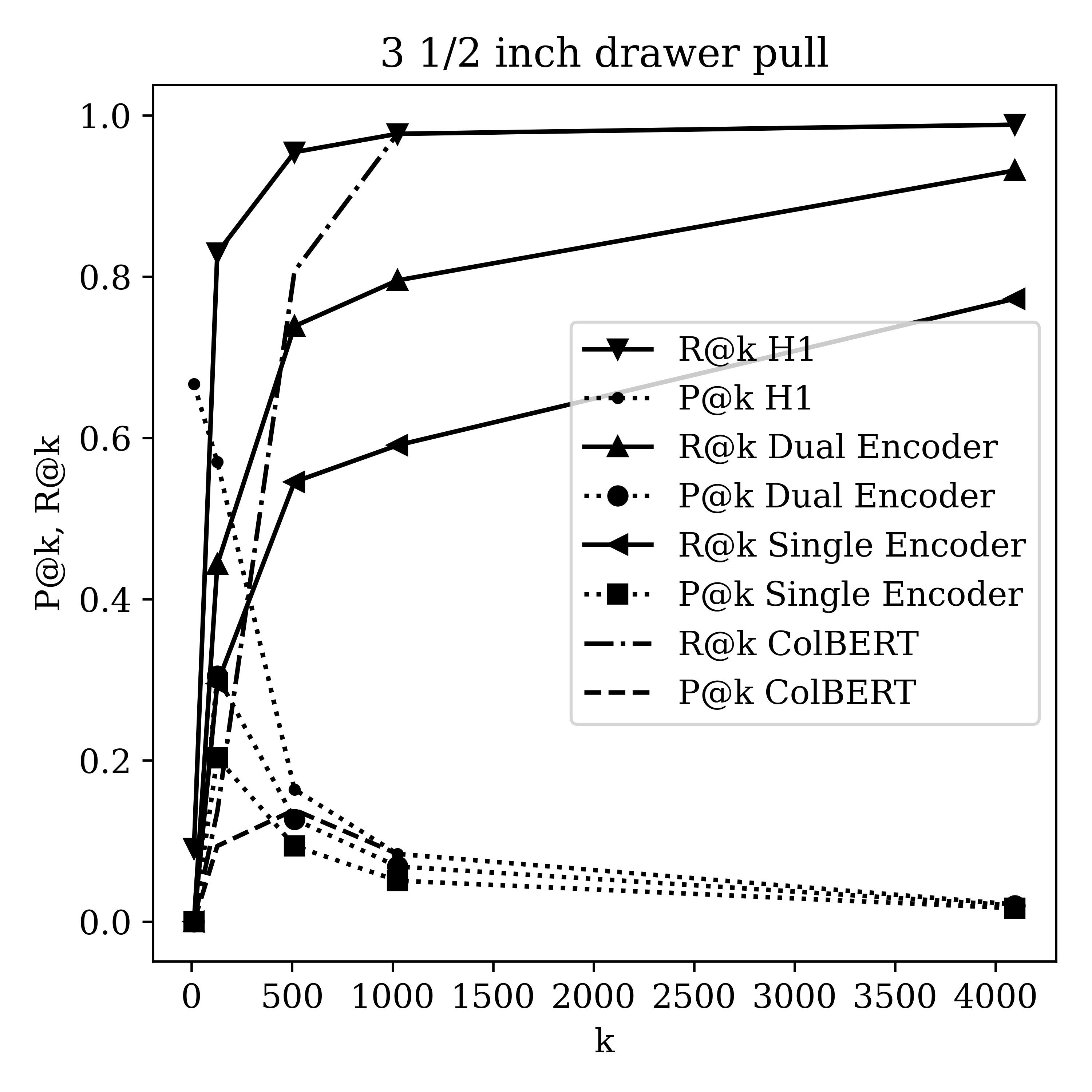}
    \caption{An illustrative one-query example of how Precision decreases and Recall increases for different semantic retrieval models with respect to cut-off threshold $k$.}
    \label{fig:single_query}
\end{figure}

The dynamics of Precision and Recall metrics for the H1 model with respect to the threshold $k$ are illustrated in Fig \ref{fig:single_query}, clearly separating the H1 model from the rest. The Recall of the search results is higher with lower values of the threshold $k$, and Precision declines more slowly as $k$ increases, compared to other models.

\section{Conclusions and Future Work}

This study introduced the H1 embedding model, a cutting-edge approach designed to refine the landscape of e-commerce search systems by leveraging multi-word term embeddings. Our extensive evaluations demonstrate that H1, through its innovative use of semantically rich tokens and hybrid search methodologies, notably enhances the accuracy and efficiency of product retrieval. By achieving mAP@12 = 56.1\% and R@1k = 86.6\% on the WANDS dataset, H1 has set a new benchmark, surpassing other state-of-the-art models in terms of precision and recall.

Our research underscores the criticality of integrating semantic understanding with traditional lexical search techniques to address the inherent limitations of each approach. The H1 model's unique ability to treat multi-word terms as singular entities not only improves the search relevance but also aligns with the natural language processing of user queries, thereby significantly enhancing the user experience in e-commerce platforms.

Future efforts will be dedicated to establishing a definitive benchmark for semantic models operating within the framework of hybrid search systems. By exploring a broader range of system architectures, the aim of our future work is to provide a comprehensive and objective evaluation framework that will not only assess the efficacy of current models but also inspire the development of more advanced and effective search solutions.

\nocite{*}
\section{References}\label{sec:reference}

\bibliographystyle{lrec-coling2024-natbib}
\bibliography{lrec-coling2024-example}

\end{document}